# Energetics of (H$_2$O)$_{20}$ isomers by means of F12 Canonical and Localized Coupled Cluster Methods

Nitai Sylvetsky[1] and Jan M.L. Martin[1, a)]

[1] *Department of Organic Chemistry, Weizmann Institute of Science, 76100 Reḥovot, Israel*

a)Corresponding author: gershom@weizmann.ac.il

**Abstract.** We consider the performance of combined PNO-F12 approaches for the interaction energies of water clusters as large as (H$_2$O)$_{20}$ by comparison to canonical CCSD(T)/CBS reference values obtained through n-body decomposition of post-MP2 corrections. We find that PNO-LCCSD(T)-F12b approaches with "Tight" cutoffs are generally capable of reproducing canonical CCSD(T) interaction energies to within ~0.25% and isomerization energies to ~1.5%, while requiring only a fraction of the canonical computational cost. However, basis set convergence patterns and effect of counterpoise corrections are more erratic than for canonical calculations, highlighting the need for canonical benchmarks on closely related systems.

## INTRODUCTION

Noncovalent interactions (NCIs) are well-known to significantly influence various physical and chemical properties of many (supra)molecular systems.[1,2] Individual NCIs (which may amount to as little as a few tenths of a kcal/mol; see, e.g., Ref.[3]), however, are extremely difficult to measure experimentally. For this reason, systematically-convergent wavefunction *ab initio* methods constitute a crucial, well-nigh exclusive source of information on such interactions. NCIs between biomolecules and water molecules, and between the latter among themselves, are essential for understanding vital biochemical mechanisms. In this context, water clusters have long been the subject of basic scientific interest because they dictate water's singular bulk properties (including but not limited to a remarkably high boiling point, low thermal expansion coefficient, and unusual density behavior). Thus, a great many experimental and theoretical studies have been carried out on the structure and properties of water in both the gas and liquid phases,[4–8] and wavefunction *ab initio* methods have already provided valuable insight into the structures and energetics of small water clusters.[9–13]

Large water clusters pose a technical challenge to wavefunction *ab initio* methodology: (H$_2$O)$_{20}$, for instance, has been studied extensively[12,14–18] for representing a transition point between the four Wales-Hodges families of three-dimensional structures (i.e., single cage, box-kite, edge-sharing- and face-sharing-prisms).[19]

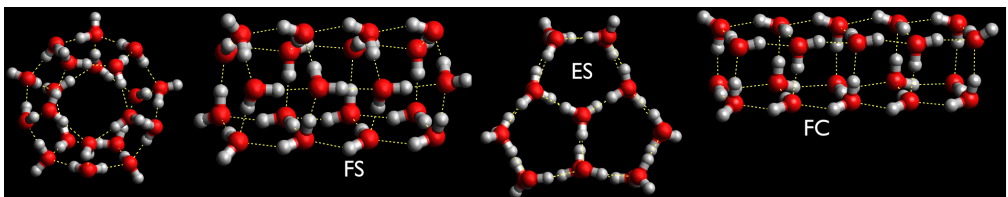

Fig. 1. Icosahedral, face-sharing, edge-sharing, and "box-kite"/face-cube structures of (H$_2$O)$_{20}$

However, canonical coupled-cluster calculations including single and double excitations as well as perturbative triples correction, i.e., CCSD(T) [or CCSD(T)-F12] calculations on systems of this size are currently beyond the reach of commodity high-performance computing (HPC) hardware, except for woefully inadequate small basis sets like aVDZ. Thus, pair natural orbital (PNO) local correlation approaches[20–23] may constitute an attractive, affordable alternative. PNO-type approaches may additionally be combined with the F12 ansatz in order to provide the best of both worlds – accelerated basis set convergence as well as nearly-linear CPU time scaling with system size.

Despite enabling *ab initio* methods to ever-larger systems of interest, practical implementations of both PNO-type and F12 approaches involve many cutoffs, screening thresholds, and other details 'under the hood'; these are hidden from the end user as tuned collections invoked through keywords like, e.g. in ORCA,[24] LoosePNO, NormalPNO (the default), and TightPNO: for details see Table 1 of Ref.[25] (The MOLPRO equivalents are specified in Tables 1-4 of Ref.[23] and the accompanying discussion.) For this reason, and especially in chemical territories where PNO-type approaches have not yet proven to be consistent and robust, comparing calculated results with a canonical benchmark is necessary to ensure the latter's reliability.

For binding energetics of water clusters, or indeed any cluster $A_n$ for large enough n, such cutoffs put an additional fly in the ointment. While they do lead to substantial pruning in $A_n$, the monomer A is so small that nothing is screened out and the PNO calculation becomes functionally equivalent to a canonical one. This creates an intrinsic imbalance for a cluster interaction energy. In addition, calculated relative energies of stable isomers for such clusters might also be affected by similar imbalances – as it might be expected that a different amount of PNOs will be screened in each isomer. In order to assess the importance of these technical difficulties, we shall examine interaction and relative energies of $(H_2O)_{20}$ clusters from the WATER27[26] dataset.

Since canonical CCSD(T) calculations in adequate basis sets are not a practical option for a reference level (or we need not resort to PNO in the first place), an alternative route must be considered. Luckily, we learned from our earlier work[27] that in an n-body expansion (see Methods section below), the high-level correction, HLC ≡ [CCSD(T) – MP2], converges fairly rapidly with *n*. Hence, a very good approximation to the canonical CCSD(T) energetics at the complete basis set (CBS) limit can be obtained from combining canonical whole-system MP2-F12/CBS with at most 4-body HLCs. Such a calculation involves thousands of single-point CCSD(T) energy calculations, but on at most tetramers, and of course it is embarrassingly parallel. This thus offers us a practical route toward nearly exact canonical benchmark results as a touchstone for the PNO approaches.

In our present analysis, we shall also consider the performance of said methods for $(H_2O)_6$ clusters – for which we were able to obtain a whole-system canonical reference level (avoiding the n-body decomposition scheme altogether). Similar error statistics on a single $(H_2O)_{10}$ cluster will also be taken into account (in this case, however, n-body decomposition was used for obtaining our reference data). Due to length limitations, our results for both $(H_2O)_6$ and $(H_2O)_{10}$ will be omitted from the present paper; we hereby refer the reader to Ref.[28]

## COMPUTATIONAL METHODS

All calculations were carried out on the Faculty of Chemistry's Linux cluster 'chemfarm' at the Weizmann Institute of Science. Single-point closed-shell PNO-LCCSD(T)-F12b[23,29] calculations were carried out using default and "domopt=tight" options, as defined in Ref.[23] and implemented in the Molpro2019 program package.[30] All PNO-based calculations discussed in this paper employ tight PNO domains unless explicitly stated otherwise.

CCSD(T)-F12b[31,32] and DF-MP2-F12[33] calculations were also used for the purpose of obtaining reliable canonical reference values. The effect of the F12b approximation was studied in detail in Refs.[34,35] and should be negligible for noncovalent interactions with VQZ-F12 and better basis sets.

For the explicitly correlated [i.e., CCSD(T)-F12b and PNO-LCCSD(T)-F12b] calculations, the correlation-consistent cc-pVnZ-F12 basis sets (a.k.a VnZ-F12) of Peterson et al.[36] were used in conjunction with the appropriate auxiliary basis sets for JKfit[37] (Coulomb and exchange), MP2fit[38,39] (density fitting in MP2), and OptRI[40,41] (complementary auxiliary basis set, CABS) basis sets. We also employed our own aug-cc-pVnZ-F12 (or aVnZ-F12) basis sets, introduced in Ref.[42]; the issue of the appropriate CABS basis set in such calculations is investigated in detail in Ref.[43] As recommended in Ref.[44], the geminal exponent (β) value was set to 1.0 for all basis sets used in explicitly-correlated calculations under consideration.

Conventional (i.e., orbital-based) *ab initio* CCSD(T)[45,46] calculations were performed using correlation-consistent[47] basis sets. In general, we used the combination of diffuse function-augmented basis sets aug-cc-pVnZ (n = T,Q,5) on non-hydrogen atoms and regular cc-pVnZ basis sets on hydrogen – to be denoted haVnZ for short.

Basis set extrapolations were carried out using the two-point formula:

$$E_\infty = E(L) - [E(L) - E(L-1)]/\left[\left(\frac{L}{L-1}\right)^\alpha - 1\right] \qquad (1)$$

where L is the highest angular momentum present in the basis set for elements B–Ne and Al–Ar and α an exponent specific to the level of theory and basis set pair. Basis set extrapolation exponents α were taken from Table 2 of Ref.[48].

Aside from Boys-Bernardi counterpoise corrections[49] and the uncorrected values, we also apply the average of both (so-called "half-CP"), as rationalized by Sherrill and coworkers[50] for orbital-based *ab initio* methods and by our group[51] for F12 calculations. In short, such practice exploits the balance between basis set superposition error (which causes overbinding) and intrinsic basis set incompleteness (which causes underbinding).[51]

As mentioned in the introduction, decomposing water cluster interaction energies (for which MP2 is a good approximation) into MP2 and HLC, and then applying an n-body expansion to the latter, may offer a route towards accurate energetics for large water clusters, where all-atom HLCs are computationally too costly.[27,52] (For brief overviews of the n-body decomposition scheme, see Refs.[53,54])

For the avoidance of doubt, all calculated interaction energies considered in this work are "vertical" – that is, the isolated monomer geometries are the same as those found within the cluster, and the interaction energy does not include monomer relaxation terms. To facilitate comparisons with earlier work, reference geometries were taken verbatim from the WATER27[26] dataset and not optimized further.

## RESULTS AND DISCUSSION

We have managed to obtain benchmark, fully-canonical results using the following level of theory:

$$\text{REF}_{water20} = \text{MP2-F12/a'V}\{T,Q\}\text{-F12 \{raw, whole system\}} + 2\text{-body}([\text{CCSD-F12b} - \text{MP2-F12}]/\text{a'VQZ-F12} + (T)/\text{AV5Z})$$
$$+ 3\text{-body}([\text{CCSD-F12b} - \text{MP2-F12}]/\text{a'VTZ-F12} + (T)/\text{a'V}\{D,T\}Z) + 4\text{-body}([\text{CCSD(T)} - \text{MP2}]/\text{a'VTZ}) \quad (3)$$

As can be seen in Ref.[28], this reference level represents an improvement over the one previously established in Ref.[27] (see ESI-2 of same paper), which corresponds to:

$$\text{REF-OLD}_{water20} = \text{MP2-F12/V}\{T,Q\}\text{-F12 \{raw, whole system\}} + 2\text{-body}([\text{CCSD-F12b} - \text{MP2-F12}]/\text{VQZ-F12}$$
$$+ (T)/\text{A'V}\{T,Q\}Z) + 3\text{-body}([\text{CCSD-F12b} - \text{MP2-F12}]/\text{VTZ-F12} + (T)/\text{a'V}\{D,T\}Z) \quad (4)$$

which neglected 4-body HLC contributions entirely. Primarily because of the latter, the new total interaction energies differ by 0.37 kcal/mol root-mean-square deviation (RMSD) from those given in Ref.[27]. Our best values are: edge-sharing pentagonal 219.19, face-sharing cubes 215.98, face-sharing pentagonal 217.03, and dodecahedron 211.58 kcal/mol, respectively. We conservatively assign an uncertainty of about 0.4 kcal/mol to these new reference values.

Error statistics for the four structures of $(H_2O)_{20}$ (Table 1) confirm our hypothesis regarding accumulation of errors in PNO-based calculations, depending on chosen PNO cutoffs – as some of the errors for the raw calculations approach 1 kcal/mol. Indeed, it can be seen in Ref.[28], that smaller such errors (below 0.86 kcal/mol) are also observed for $(H_2O)_{10}$. For $(H_2O)_6$, however, they are virtually nonexistent. Unfortunately, mean signed deviations (MSD) for PNO-based methods are not consistently underbound nor overbound [as in the case of $(H_2O)_{10}$] – which precludes using an *ad hoc*, *a posteriori* correction for PNO-based results – such as a constant scaling factor – in order to eliminate biases from corresponding canonical limits.

|  |  | RMSD |  |  |  |  | MSD |  |  |  |  |
|---|---|---|---|---|---|---|---|---|---|---|---|
|  | n= | T | T* | Q | Q* | {T/Q} | T | T* | Q | Q* | {T/Q} |
| RAW | VnZ-F12 | 0.309 | 0.430 | 0.504 | 0.225 | 0.533 | -0.243 | +0.398 | -0.494 | -0.209 | -0.531 |
|  | a'VnZ-F12 | 0.511 | 0.271 | 0.942 | 0.605 | 1.092 | -0.498 | +0.258 | -0.940 | -0.603 | -1.091 |
| CP | VnZ-F12 | 2.556 | 1.408 | 1.323 | 0.756 | 0.641 | -2.553 | -1.406 | -1.322 | -0.755 | -0.641 |
|  | a'VnZ-F12 | 1.938 | 0.764 | 1.190 | 0.619 | 0.753 | -1.935 | -0.760 | -1.190 | -0.620 | -0.752 |
| Half-CP | VnZ-F12 | 1.407 | 0.518 | 0.911 | 0.485 | 0.586 | -1.398 | -0.504 | -0.908 | -0.482 | -0.586 |
|  | a'VnZ-F12 | 1.221 | 0.261 | 1.065 | 0.612 | 0.922 | -1.216 | -0.250 | -1.060 | -0.610 | -0.922 |

* = Constant scaling of triples (Ts), cf. Table 3 of Ref.[55]

**Table 1.** $(H_2O)_{20}$: Error statistics (kcal/mol) for PNO-LCCSD(T)-F12b interaction energies obtained using various basis sets; $\text{Ref}_{water20}$, Eq. (3), is used as a reference.

Nevertheless, PNO-based methods can still be used to reproduce canonical reference values for $(H_2O)_{20}$ to within 0.5 kcal/mol, which at about 0.25% of the cluster association energy might be deemed negligible in relative terms. As for $(H_2O)_6$ and $(H_2O)_{10}$, CP corrections prove to be quite ineffectual, and do not justify the required additional computational cost (i.e., using the entire cluster's basis functions for calculating monomer energies). (Ts) scaling (constant scaling by a fixed ratio determined for a small training set – given in Table 3 of Ref.[55]), on the other hand, does seem to improve most calculated results – as even a′VTZ-F12, which is relatively compact and computationally economical, comes close to the reference. Thus, a PNO-LCCSD(T)-F12b calculation with tight PNO settings on the edge-sharing $(H_2O)_{20}$ structure took 6 days and 14 hours CPU time (with nearly perfect parallelization), whereas our n-body-based canonical reference calculation according to eq. (3) required more than a year (!) for the same system. This is indeed good news, as PNO-based methods can now be recognized as both remarkably economical and fairly accurate for the systems under consideration.

Would switching to default PNO domains come at a substantial cost in accuracy? As can be seen in Table 2, employing such settings leads to further under-binding: for the a′VnZ-F12 (n=T,Q) basis sets considered here, applying default PNO domains results in an error two to six times as large than that obtained using tight PNO settings. That being said, it should once more be noted that default PNO settings are more computationally economical, and may thus be preferable for larger systems for which tight PNOs are too demanding. Again, a PNO-LCCSD(T)-F12b calculation with tight PNO settings on the edge-sharing $(H_2O)_{20}$ structure required 6 days and 14 hours CPU time, compared to only 23 hours with default PNOs (6.843:1 ratio). This finding may indeed be useful for, say, still larger water clusters [such as $(H_2O)_{100}$]. However, the associated compromises on accuracy are not quite justified for the case under consideration.

|  | RMSD Raw | | | | | MSD Raw | | | | |
|---|---|---|---|---|---|---|---|---|---|---|
| a'VnZ-F12 | T | T* | Q | Q* | {T/Q} | T | T* | Q | Q* | {T/Q} |
| tightDomain | 0.361 | 0.191 | 0.666 | 0.428 | 0.772 | -0.498 | 0.258 | -0.940 | -0.603 | -1.091 |
| default | 1.689 | 1.203 | 1.716 | 1.494 | 1.641 | -2.381 | -1.694 | -2.421 | -2.107 | -2.316 |

\* = Constant scaling of triples (Ts), cf. Table 3 of Ref.[55]

**Table 2.** $(H_2O)_{20}$: Error statistics (kcal/mol) for PNO-LCCSD(T)-F12b interaction energies obtained using a'VnZ-F12 (n=T,Q) basis sets with tight and default PNO domains; Ref$_{water20}$, eq. (3), is used as a reference.

Would various truncation errors in the PNO result inject a random, rather than systematic, error component into *relative* energies of the different structural isomers? This would be a major downside of PNO methods if true, since for canonical approaches, relative energies of structures are well known to converge much faster and more smoothly with basis set and electron correlation approach than total interaction energies. As can be seen in Table 3 for the relative energies of the four $(H_2O)_{20}$ isomers, however, PNO-LCCSD(T)-F12b methods clearly reproduce canonical reference values without compromising much on accuracy: raw PNO-LCCSD(T)-F12b/a'VTZ-F12 with default PNO domains underestimates the icos–ES difference by ~0.4 kcal/mol (or by ~5%), but makes significantly smaller errors for the FC–ES and FS–ES energetic gaps (0.05 and 0.02 kcal/mol, or ~1% and ~1.6% respectively). Switching to tight PNO domains and larger basis sets further reduces these errors – which amount to just 0.01 (icos–ES), 0.03 (FC–ES), and 0.09 (FS-ES) kcal/mol using raw PNO-LCCSD(T)-F12b/a'V{T,Q}Z-F12 with tight PNO domains. Thus, we see that errors in relative energies obtained using PNO methods converge relatively smoothly to the canonical basis set limit – as previously observed for calculated interaction energies.

| a'VnZ-F12 | T | T* | Q | Q* | {T/Q} | Ref. Values | T | T* | Q | Q* | {T/Q} |
|---|---|---|---|---|---|---|---|---|---|---|---|
|  | tightDomain | | | | | | defaultDomain | | | | |
| FC – ES | 0.039 | 0.025 | 0.027 | 0.020 | 0.015 | **3.204** | 0.052 | 0.041 | -0.043 | -0.049 | -0.096 |
| FS – ES | -0.056 | -0.057 | -0.037 | -0.037 | -0.032 | **2.151** | -0.021 | -0.020 | -0.089 | -0.090 | -0.126 |
| icos – ES | -0.260 | -0.187 | -0.146 | -0.113 | -0.093 | **7.608** | -0.422 | -0.360 | -0.414 | -0.384 | -0.378 |

**Table 3.** $(H_2O)_{20}$: Signed deviations (kcal/mol) for PNO-LCCSD(T)-F12b relative isomer energies obtained using a'VnZ-F12 (n=T,Q) basis sets with tight and default PNO domains; canonical reference values are also given (upper right; Ref$_{water20}$, eq. (3), is used as the reference level). Interaction energy for the edge-shared global minimum is kcal/mol.

To sum up: we have seen that PNO-based methods can be used to reproduce accurate canonical interaction energies of water clusters up to $(H_2O)_{20}$. This further corroborates that such methods are a viable alternative to canonical calculations for calculating NCIs of biologically relevant systems. That being said, and since systematic convergence is perhaps *the* key benefit of wavefunction *ab initio* calculations, the somewhat erratic behavior of PNO-based regarding CP corrections and basis set extrapolations is indeed somewhat troubling. Thus, we can again see that results obtained using such methods should be treated with some caution in cases where no calibration against canonical values is available.

## ACKNOWLEDGMENTS

Research at Weizmann was funded by the Israel Science Foundation (grant 1358/15) and by the Estate of Emile Mimran. NS acknowledges a doctoral fellowship from the Feinberg Graduate School at the Weizmann Institute. The authors would like to thank Profs. Frank Neese and Hans-Joachim Werner for helpful discussions.

# Energetics of (H$_2$O)$_{20}$ isomers by means of F12 Canonical and Localized Coupled Cluster Methods

*Nitai Sylvetsky[§] and Jan M. L. Martin[*,§]*

Electronic Supporting Information

[§] Department of Organic Chemistry, Weizmann Institute of Science, 76100 Reḥovot, Israel.

Email: gershom@weizmann.ac.il FAX: +972 8 934-3029

| System | Interaction Energy (kcal/mol) |
|---|---|
| 4192_water6BAG | **47.281** |
| 4193_water6BK1 | **48.015** |
| 4194_water6BK2 | **47.741** |
| 4195_water6CA | **48.437** |
| 4196_water6CB1 | **45.907** |
| 4197_water6CB2 | **45.807** |
| 4198_water6CC | **46.911** |
| 4199_water6PR | **48.776** |

**Table S1.** Best estimates for $(H_2O)_6$ interaction energies (reference level: CCSD(T)/A'V{Q,5}Z, Half-CP)

**Root-Mean-Square Deviation**

**Raw**

| n= | T | T* | Q | Q* | 5 | {T/Q} | {Q/5} |
|---|---|---|---|---|---|---|---|
| VnZ-F12 | 0.102 | 0.228 | 0.063 | 0.024 | - | 0.109 | - |
| a'VnZ-F12 | 0.024 | 0.313 | 0.173 | 0.021 | - | 0.240 | - |
| conv. haVnZ | 0.099 | - | 0.030 | - | 0.150 | 0.032 | 0.249 |

**Counterpoise**

| n= | T | T* | Q | Q* | 5 | {T/Q} | {Q/5} |
|---|---|---|---|---|---|---|---|
| VnZ-F12 | 0.507 | 0.389 | 0.242 | 0.187 | - | 0.164 | - |
| a'VnZ-F12 | 0.390 | - | 0.206 | - | - | 0.140 | - |
| conv. haVnZ | 3.152 | - | 1.188 | - | 0.609 | 0.247 | 0.139 |

**half-Counterpoise**

| n= | T | T* | Q | Q* | 5 | {T/Q} | {Q/5} |
|---|---|---|---|---|---|---|---|
| VnZ-F12 | 0.206 | 0.097 | 0.151 | 0.095 | - | 0.135 | - |
| a'VnZ-F12 | 0.189 | - | 0.189 | - | - | 0.189 | - |
| conv. haVnZ | 1.615 | - | 0.608 | - | 0.379 | 0.130 | 0.194 |

**Mean Signed Deviation**

**Raw**

| n= | T | T* | Q | Q* | 5 | {T/Q} | {Q/5} |
|---|---|---|---|---|---|---|---|
| VnZ-F12 | 0.098 | 0.227 | -0.057 | 0.000 | - | -0.104 | - |
| a'VnZ-F12 | 0.013 | 0.313 | -0.171 | 0.019 | - | -0.238 | - |
| conv. haVnZ | -0.080 | - | -0.027 | - | -0.149 | 0.011 | -0.247 |

**Counterpoise**

| n= | T | T* | Q | Q* | 5 | {T/Q} | {Q/5} |
|---|---|---|---|---|---|---|---|
| VnZ-F12 | -0.506 | -0.388 | -0.187 | -0.187 | - | -0.163 | - |
| a'VnZ-F12 | -0.389 | - | -0.206 | - | - | -0.139 | - |
| conv. haVnZ | -3.149 | - | -1.186 | - | -0.608 | 0.247 | -0.139 |

**half-Counterpoise**

| n= | T | T* | Q | Q* | 5 | {T/Q} | {Q/5} |
|---|---|---|---|---|---|---|---|
| VnZ-F12 | -0.204 | -0.095 | -0.150 | -0.094 | - | -0.133 | - |
| a'VnZ-F12 | -0.188 | - | -0.188 | - | - | -0.188 | - |
| conv. haVnZ | -1.615 | - | -0.607 | - | -0.378 | 0.129 | -0.193 |

\* = Constant scaling of triples (Ts), cf. Table 3 of K.A. Peterson, M.K. Kesharwani, J.M.L. Martin, *The cc-pV5Z-F12 basis set: reaching the basis set limit in explicitly correlated calculations*, Mol. Phys. 113 (2015) 1551–1558.

**Table S2.** Eight isomers of $(H_2O)_6$: Error statistics (kcal/mol) for PNO-LCCSD(T)-F12b interaction energies obtained using various basis sets; half-CP corrected, canonical CCSD(T)/A'V{Q,5}Z is used as a reference.

| | raw | | | | |
|---|---|---|---|---|---|
| n= | T | T* | Q | Q* | {T/Q} |
| VnZ-F12 | 0.561 | 0.856 | 0.412 | 0.543 | 0.304 |
| a'VnZ-F12 | 0.409 | 0.757 | 0.204 | 0.358 | 0.054 |
| | CP | | | | |
| n= | T | T* | Q | Q* | {T/Q} |
| VnZ-F12 | -0.505 | -0.233 | 0.035 | 0.162 | 0.429 |
| a'VnZ-F12 | -0.273 | 0.053 | 0.094 | 0.245 | 0.362 |
| | half-CP | | | | |
| n= | T | T* | Q | Q* | {T/Q} |
| VnZ-F12 | 0.028 | 0.311 | 0.224 | 0.353 | 0.367 |
| a'VnZ-F12 | 0.068 | 0.405 | 0.149 | 0.302 | 0.208 |

* = Constant scaling of triples (Ts), cf. Table 3 of Ref.[67]

**Table S3.** $(H_2O)_{10}$: Signed deviation (kcal/mol) for PNO-LCCSD(T)-F12b interaction energies obtained using various basis sets (reference interaction energy: **98.504** ; reference level: whole-system MP2-F12/a'V{T,Q}Z-F12, half-CP + 2-body[CCSD(T) – MP2]/AV{Q,5}Z + 3-body[CCSD(T) – MP2]/AV{T,Q}Z + 4-through-10-body[CCSD(T) – MP2]/AVTZ).